\documentclass[12pt]{iopart}
\usepackage{iopams} 
\usepackage{amssymb}
\usepackage{amsfonts}
\usepackage{graphicx}
\usepackage{helvet}
\usepackage{hyperref}
\usepackage{caption}
\usepackage{isotope}
\usepackage{xfrac}
\usepackage{cite}
\usepackage{color}
\usepackage{tikz}
\usetikzlibrary{arrows.meta}
\tikzset{%
  >={Latex[width=2mm,length=2mm]},
            base/.style = {rectangle, rounded corners, draw=black,
                           minimum width=4cm, minimum height=1cm,
                           text centered, font=\sffamily},
  neutron/.style = {base, fill=blue!80},
       beta/.style = {base, fill=blue!20},
    stable/.style = {base, fill=black!40},
         alfa/.style = {base,  fill=yellow,
                           font=\ttfamily},
 }                          

\begin{document}

\title[From nuclei to neutron stars (part 2)]{From nuclei to neutron stars: simple binding energy computer modelling in the classroom (part 2)}

\author{ A Rios }
\address{Department of Physics, Faculty of Engineering and Physical Sciences, University of Surrey, Guildford, Surrey GU2 7XH, United Kingdom}
\address{
Departament de F\'isica Qu\`antica i Astrof\'isica, 
 Institut de Ci\`encies del Cosmos (ICCUB),
Universitat de Barcelona, Mart\'i i Franqu\`es 1, E08028 Barcelona, Spain
}
\ead{a.rios@surrey.ac.uk}

\author{ A Pastore }
\address{Department of Physics, University of York, Heslington, York, Y010 5DD, United Kingdom }
\ead{alessandro.pastore@york.ac.uk}

\author{ C Diget }
\address{Department of Physics, University of York, Heslington, York, Y010 5DD, United Kingdom }

\author{ A M Romero}
\address{Department of Physics and Astronomy, The University of North Carolina at Chapel Hill, Chapel Hill, North Carolina 27599, USA}

\author{ K Leech }
\address{Department of Physics, University of York, Heslington, York, Y010 5DD, United Kingdom }

\author{ P Stokoe }
\address{Department of Physics, University of York, Heslington, York, Y010 5DD, United Kingdom }

\begin{abstract}
We introduce two simple online activities to explore the physics of neutron stars. These provide an introduction to the basic properties of compact objects, like their masses and radii, for secondary school students. The first activity explores the idea of the minimum mass of a neutron star. It is directly linked to the concept of binding energy and follows on from our previous activities. The second activity focuses on the maximum mass of neutron stars using a solvable model of the neutron star interior.
The activities are based on spreadsheets, provided as Supplementary Material,  and can be easily adapted to different levels, age groups and discussion topics. In particular, these activities can naturally lead towards discussions on extrapolations and limits of theoretical models. 
\end{abstract}

\section{Introduction}

The idea behind the \emph{Binding Blocks} (BB) project is to explore physics outreach in an interactive way, employing physical LEGO\textsuperscript{\textregistered} bricks\footnote{LEGO\textsuperscript{\textregistered} is a trademark of the LEGO Group of companies which does not sponsor, authorise or endorse the present work.} to provide a clear understanding of key physics concepts. In a typical BB setting, the audience is asked to create a section of the nuclide chart, with towers representing different isotopes. In doing this, the audience actively explores ideas surrounding elements and isotopes. Using different colours, one can also identify nuclear decays, and the height of the towers can illustrate, for instance, the binding energy  \cite{BB17}. All in all, the activities promote an understanding of key ideas in a wide range of physics contexts - including isotopic sciences, nuclear models and astrophysics. 

The BB activities are amenable to different types of audiences, although the focus is on secondary school A-level students for whom specific activities have been created~\cite{Wright2017}. A series of additional outreach activities have evolved around the BB effort, complementing the physical LEGO\textsuperscript{\textregistered}  nuclide chart. A freely available online 3D chart, for instance, provides a visualization of the same ideas without having to physically access LEGO\textsuperscript{\textregistered} bricks~\cite{Simpson2017}. The BB programme also has a YouTube\textsuperscript{\textregistered} channel, where short lectures are regularly broadcast. Complementing these online efforts, the University of York successfully ran online Nuclear Masterclasses in 2020, attracting over 2000 participants~\cite{Masterclass2020}. 

In a previous paper (hereafter referred to as Part 1~\cite{Part1_2020}), we have introduced a series of online, spreadsheet-based activities that provide an introduction to nuclear physics models in the context of nuclear binding. Unlike previous BB efforts based on experimental nuclear data~\cite{BB17} or on nuclear processes~\cite{frost2017cycling}, these activities focus on a \emph{theoretical} model. Using a simple mathematical formula based on the liquid-drop (LD) model, students engage with hands-on activities in an online setting. The experience is embedded in an active learning environment that allows for the exploration of three key issues of theoretical modelling: optimisation, validation and prediction. 

The spreadsheet materials for these activities provide an immediate visualization, offering a direct understanding these details without any cumbersome derivation. A (simple) mathematical model is introduced to provide context and physical guidance.
To optimise, students are asked to fit the simple LD formula to a series of nuclear binding energy data. The validation step is addressed qualitatively by visually analysing the discrepancies between the model predictions and data. Finally, students are asked to predict the masses of the largest bound isotopes in a given chain. With this, in addition to providing an insight into nuclear binding, we show that theory models can describe physical data; that optimisation is often employed in theory; and that mathematical models can be used to predict physical phenomena. If instructors feel they are of interest, these spreadsheets can also be used to explore issues around statistical and systematic errors for more advanced students. The activities can also be used to discuss the role of modelling in science; its advantages and its potential limitations.

All in all, the activities we have presented provide a complete and consistent set that can be used to illustrate the key stages of scientific modelling \cite{hickman1986mathematical,Gilbert2004}. When it comes to the model prediction stage, however, one often makes a distinction between interpolation and extrapolation in terms of the domain of the model. When the model is used to predict data within its initial domain, it is often said to \emph{interpolate}. When, in contrast, the model is exploited beyond its initial remit, one is generally \emph{extrapolating}. Extrapolation based on models has lead to discoveries across the scientific domain, from the discovery of the helical structure of DNA \cite{Franklin1953,Watson1953,Schindler2008} to aspects of exoplanet research \cite{Fressin2013}.

Here, we create a series of student activities that can help explore issues around model extrapolation, and its uses in the scientific endeavour. 
In particular, we extend the ideas of Part 1 and provide two online activities that focus not on the physics of atomic nuclei, but rather on the astrophysics of compact objects. In particular, we discuss the properties of an altogether different system: neutron stars (NSs)~\cite{shapiro1983black,Haensel}. This is relevant for a variety of reasons. First, because the theoretical model allows to bridge the gap between physical systems which are 19 orders of magnitude apart in size. This is something that can not be done with experimental data alone. Second, because it provides a connection to astrophysics, a field that is not necessarily considered ``nuclear" in nature. This illustrates the multidisciplinary applications of scientific models. Finally, there is a relative scarcity of resources on neutron-star physics for A-level teachers and students, in stark contrast to the undergraduate and graduate levels \cite{Balian1999,Silbar2004,Jackson2005,Haensel,shapiro1983black}. This is a first attempt to provide activities that illustrate the extreme properties of these stars, the densest known objects in equilibrium, for secondary students. We feel that this contribution is particularly timely in view of the recent discovery of binary neutron stars with gravitational waves~\cite{GW170817,GW190425,boyle2019teaching}. 

In our first proposed activity, we extend the LD model of Part 1~\cite{Part1_2020} to the neutron-rich regime and supplement it with a binding energy term of a gravitational origin. By looking at the limits of binding in this case, we can derive an expression for the \emph{minimum} mass of a neutron star~\cite{Heyde_book}. 
The model can be easily derived with pen and paper, and could be used as a kick-off point to discuss the extraordinary properties of neutron stars. In addition, we provide a Worksheet to achieve a visual and interactive discussion. 

The second activity focuses, in contrast, on the concept of a \emph{maximum} mass for neutron stars. This is akin to the idea of the maximum mass of white dwarfs - the well-known Chandrasekhar limit \cite{Chandrasekhar1931,Chandrasekhar1935,Jackson2005}. 
The Chandrasekhar mass limit is a direct consequence of the lack of gravitational stability associated to relativistic fluids \cite{Kippenhahn_book,Hansen_book}. In contrast, the origin of the maximum mass limit of NSs (often referred to as the Tolman-Oppenheimer-Volkoff limit) is more complex and involves, at least partially, general relativity~\cite{Haensel,Misner}. Our activity does not require knowledge of any of these advanced undergraduate concepts, but rather relies on a known analytical solution for the neutron-star interior. We use this simple model to illustrate the extreme properties of compact objects in terms of density and pressure. While some numerical values are linked to the binding energy ideas, this activity is disconnected from any nuclear physics background and provides, instead, an astrophysical analysis. Removing the nuclear physics aspects allows for a full use of \emph{extrapolation} beyond the original remit of the theory. Furthermore, the appearance of this maximum mass is the consequence of a limit within the analytical solution. In other words, this model predicts its own breakdown which, in itself, is a surprising feature. 

The two proposed activities can be run in different ways. They can, for instance, be used as online individual activities for independent students. A minimal instruction set and some online background material (such as the video here\footnote{https://www.youtube.com/watch?v=Qsu7IrGiOIk}) may be helpful in this case.  Alternatively, the activities could be part of a group discussion or a demonstrator-led outreach or educational session - in conjunction with those presented in Part 1~\cite{Part1_2020}. Either way, the spreadsheets provided can be used to illustrate visually concepts associated to binding, nuclear physics, astrophysics and general relativity. 

The article is organised as follows. We start with a generic discussion on binding energies and the LD model in the following section, providing a short overview of Part 1~\cite{Part1_2020}. Section~\ref{sec:star} is devoted to the neutron-star activities. The minimum mass activity, following on from our previous work,  is discussed in depth in Subsection~\ref{subsec:min}. We discuss some initial astrophysical considerations of neutron stars in Subsection~\ref{subsec:astro}. Subsection~\ref{subsec:max} is instead devoted to an activity on neutron star maximum masses. 
We discuss the key learning outcomes of these activities and provide conclusions in Sec.~\ref{sec:concl}.

%%%%%%%%%%%%%%%%%%%%%%%%%%%

\section{The Liquid Drop model}\label{sec:LD}

A composite physical systems is expected to be bound when it is energetically favourable to keep the system together. The amount of energy that is required to separate the system in all its independent constituents is the \emph{binding energy}. For an isotope of element $Z$ with $N$ neutrons, the binding energy is given by the difference
\begin{eqnarray}\label{eq:BE}
BE&=N m_n c^2 + Z m_p c^2 - M_{N,Z} c^2 \, .
\end{eqnarray}
Here, $m_nc^2= 939.565$ MeV\footnote{In nuclear physics activities, it is more natural to work in units of mega-electronvolts (MeV)  rather than Joules (J). We recall that  $1$ J=$6.242\cdot 10^{12}$ MeV} and $m_pc^2=938.272$ MeV represent the rest mass energies of the neutron and the proton, respectively. $M_{N,Z}c^2$ is the rest mass of the isotope. More details and introductory activities on binding energies are available in Part 1~\cite{Part1_2020} and Ref.~\cite{baroni2018teaching}.

The typical binding energies of medium-to-heavy nuclei are of the order of $\approx 8$ MeV per particle (nucleon) in the nucleus. Nuclear masses, $M_{N,Z}c^2$, can be measured to exquisite accuracy. Out of the $\approx 3400$ isotopes that have been identified experimentally so far, more than $2000$ binding energies per particle have been measured with accuracies well below a fraction of a percent~\cite{wang2017ame2016}. These masses are relevant for physics applications beyond pure nuclear science. For instance, the nucleosynthesis of isotopes beyond mass $A=56$~\cite{koura2014three} is largely due to the so-called rapid neutron capture process (or r-process) - which is sensitive to nuclear masses.

%%%%%%%%%%%%%%

\begin{table}[!t]
\begin{center}
\begin{tabular}{|c|c|c|}
\hline
\hline
 \multicolumn{2}{|c|}{Coefficients [MeV]}\\
\hline
$a_V$& 15.8   \\
$a_S$&   18.3  \\
$a_C$& 0.714  \\
$a_{A}$& 23.2   \\
$a_{P}$& 12.0  \\
\hline
\hline
\end{tabular}
\end{center}
\caption{Coefficients (expressed in MeV) of the LD mass formula Eq.~(\ref{eq:LD}). }
\label{tab:coeff}
\end{table}

From a theoretical perspective, it is difficult to provide a microscopic description of nuclear masses with the same level of accuracy that is reached in experiments \cite{Shelley2021,Wu2020}. We can, however, devise physical models that capture the key physics that is relevant to determine nuclear masses. This is precisely what the LD model achieves, starting from a set of relatively simple physical concepts and providing a mathematical expression that is widely used - the so-called Bethe-Weizs\"acker mass formula~\cite{hodgson1997introductory}. This mathematical model provides the binding energy per particle of an isotope with proton number $Z$, neutron number $N$, and mass number $A=N+Z$:
\begin{eqnarray}\label{eq:LD}
\frac{BE}{A}&=a_V-\frac{a_S}{A^{1/3}}-a_C\frac{Z^2}{A^{4/3}}-a_{A}\frac{(N-Z)^2}{A^2}+a_P\frac{\delta_{N,Z}}{A} \, .
\end{eqnarray}
We take the values for the coefficients $a_X$ from Ref.~\cite{Weiz1935} and quote them for completeness in Table~\ref{tab:coeff}. 
We refer the reader to Part 1~\cite{Part1_2020} for more details on how this formula is derived, including the definition of the pairing function $\delta_{N,Z}$.

In Part 1~\cite{Part1_2020}, we provided a series of activities that exploit Eq.~(\ref{eq:LD}) and simple spreadsheet visualization tools to provide an immediate understanding of binding energies. The three suggested activities deal with different key stages that are relevant not only in nuclear physics, but for theoretical models in general. In a first stage, the model is \emph{optimized}, in the sense that the coefficients are determined to match a given subset of data. In a second stage, the model is \emph{validated}: a process that typically involves discussing quality measures with respect to the original subset of data. The validation process can also be extended to datasets that are close in some sense to the original one. Finally, the third stage involves a \emph{prediction}. The model is used not to discuss known data, but to provide a way forward towards an unknown domain. 

In the scientific process the final step, prediction, is key. In fact, scientific models are often characterised by their ability to provide predictions that  either validate the model or discard it. 
Mathematical formulae, however, can also be used to reach beyond the initial remit of the physical model where they were initially formulated. This is what we call \emph{extrapolation}~\cite{Pastore2021}. In the following, we will use the LD model and extrapolate its applicability from nuclei to an astrophysical compact object, neutron stars.

\section{Neutron stars}
\label{sec:star}

Having worked with the LD model in an optimisation, validation and prediction settings, at this stage the students should have a good idea of the typical order of magnitude of nuclear sizes; how many nucleons are expected to be bound in a nucleus, and what are the largest nuclei that may be found using LD formula. The aim of this two additional activities is twofold. First, the students will use the LD model to perform an extreme extrapolation that challenges, using mathematical means, some of the preconceived ideas in Part 1. This extrapolation activity may be counter-intuitive at first, but remains entirely within the domain of scientific exploration. 

Second, we follow our first activity with another one that is fully astrophysical in nature. The data used in this second activity depends on the results of the first and provides an insight on the extreme interior properties of neutron stars. This two-step procedure illustrates the power of scientific extrapolation. In this context, the process can also be an illustration as to why nuclear physics knowledge is not only auxiliary, but  necessary, in the physical understanding of compact stellar objects. 

\subsection{Minimum mass}
\label{subsec:min}

At this stage, it may be important for instructors to recall the \emph{fundamental forces} acting in a nucleus. So far, the LD model that we introduced in Part 1~\cite{Part1_2020} and summarised in Eq.~(\ref{eq:LD}), only considers the strong nuclear force among protons and neutrons as well as the electromagnetic interaction among protons. The demonstrator may now ask why the gravitational attraction between nucleons was not taken into account. This question can be answered by looking at the relative strength of the electromagnetic and the gravitational forces in a nucleus. On the one hand, one can evaluate the strength of the electromagnetic force, $F_{ch}$, acting between two protons of charge $e_p$ within a nucleus. On the other, one can estimate the corresponding gravitational attraction $F_g$ between two protons. For these two estimates, one can use SI units and a typical nuclear distance scale of $d \approx 10^{-15} \;\;\text{m}$. With this in mind, we obtain
\begin{eqnarray}\label{eq:chg}
F_{C}=-\frac{1}{4 \pi \epsilon_0} \frac{e_p^2}{d^2}\approx -200 \;\; \text{N}\;, \label{eq:Fch} \\
F_g=G\frac{m_p^2}{d^2}\approx10^{-34} \;\;\text{N}\;,
\end{eqnarray}
\noindent where we have used $\epsilon_0=8.85 \cdot 10^{-12} \;\; \text{F m}^{-1} \text{ C}^{-2}$; $G=6.67 \times 10^{-11}\text{N m}^2 \text{ kg}^{-2}$; the mass of a proton, $m_p=1.67 \times 10^{-27}$ kg; and its charge $e_p=1.60 \times 10^{-19} \;\;\text{C}$~\cite{constants}.
The large difference, of $36$ order of magnitude, clearly illustrates that gravitational effects can be safely neglected in typical nuclear properties. It is worth pointing out that, even though we do not tackle it explicitly here, the nuclear force acting between nucleons is even larger than the electromagnetic estimate of Eq.~(\ref{eq:Fch}). 

An equivalent measure on the (ir)relevance of gravitational effects can be obtained not from forces, but from energies. This different illustration serves the same purpose as the force example above, but is worked out in typical nuclear units (rather than SI units). This change may help students familiarise themselves with a different set of units. As we have seen, typical nuclear energies (such as, say, binding energies per particle) are of the order of $1-10$ MeV. In contrast, we can calculate the gravitational energy $E_g$ for a heavy stable isotope like $^{208}$Pb by means of the formula
\begin{eqnarray}\label{grav}
E_g=\frac{3}{5}\frac{G M_{^{208}\mathrm{Pb}}^2}{R_{^{208}\mathrm{Pb}}} 
\approx 4\cdot 10^{-33}\;\text{MeV}\;.
\end{eqnarray}
\noindent 
Here, we have assumed that the interior of the nucleus can be modelled as a constant density sphere, which explains the factor $\frac{3}{5}$ (but any other constant of order $1$ would do). For this estimate, we also employed the values in typical nuclear units provided in Table~\ref{tab:conv}, with an approximate nuclear mass $M_{^{208}\mathrm{Pb}} c^2 \approx 208 \times m_p c^2 $ and a nuclear radius $R_{^{208}\mathrm{Pb}}=7.11$ fm from the standard nuclear formula
\begin{equation}
\label{eq:radius}
R=r_0 A^{1/3} \, ,
\end{equation}
with $r_0=1.2$ fm \cite{Heyde_book,hodgson1997introductory}.

\begin{table}
\begin{center}
\begin{tabular}{|c|c|c|}
\hline
 \multicolumn{1}{|c|}{}
&SI & Nuclear units\\
\hline
$r_0 $ & $1.2\cdot10^{-15}$ m & 1.2 fm\\
$\hbar c$ & $3.15\cdot 10^{-26}$ J m & 197.3 MeV fm\\
$m_n$ &  $1.67 \cdot 10^{-27}$ kg& 939 MeV c$^{-2}$ \\
$\phantom{a}$ & $G=6.67\cdot 10^{-11}$ Nm$^2$ kg$^{-2}$ & $\frac{G}{\hbar c}=6.708\cdot10^{-45}$ c$^{4}$ MeV$^{-2}$\\
\hline
\hline
\end{tabular}
\end{center}
\caption{Table of conversion between SI units and nuclear units. }
\label{tab:conv}
\end{table}

The demonstrator can stress that the contribution of Eq.~(\ref{grav}) is negligible in finite (normal) nuclei, due to the very small value of the gravitational constant $G$. The only way in which one could obtain a non-negligible contribution from Eq.~(\ref{grav}) would be to enormously increase the number of nucleons \cite{Heyde_book} - possibly to a macroscopic number. We can use the LD model to extrapolate to a system where this is possible. 
Let us first assume that such a system exists and is bound. We refer to the mass and radius of a macroscopic, bound nuclear object as $M_S$ and $R_S$. By inspecting Eq.~(\ref{eq:LD}), we observe that the Coulomb repulsion grows as $Z^2$ and decreases the binding energy. As a consequence, we may look for objects with very small number of protons and, for simplicity, we  take $Z=0$. Notice that since $Z=0$, we can use the identity $A=N$. Under these conditions, we find that the LD formula becomes
\begin{eqnarray}
\frac{BE}{N}=a_{V}-\frac{a_{S}}{N^{1/3}}-a_{A} + a_P \frac{\delta}{N} \, .%+ \frac{3 G M_S^2}{5 N R_S} \, ,
\end{eqnarray}
For a very large system, $N \gg 1$, and the terms proportional to $a_S$ and $a_P$ can be safely discarded. 

In other words, the nuclear contribution to the binding of a macroscopic neutron-rich object is $\frac{BE}{N} \approx a_{V}-a_{A}$. This (nuclear) term is independent of $N$ and, when summed up, the two constants provide an overall negative value ($-7.4$ MeV for the values in Table~\ref{tab:coeff}). With the standard sign definition of binding energies, this indicates that the system would not be bound by the strong force. For a macroscopic system, however, gravity may provide enough binding energy to change the sign of $BE$. 
For the sake of discussion, we approximate the total gravitational mass of this object by the mass of all its constituent neutrons, $M_S=N \times m_n$. We estimate its radius using the standard formula, Eq.~(\ref{eq:radius}), replacing $A \to N$, so that $R_S = r_0 N^{1/3}$.
The corresponding gravitational energy is therefore $\frac{3 G M_S^2}{5 N R_S} =  \frac{3 G m_n^2}{5  r_0} N^{2/3 }$, and we add this attractive contribution to the (repulsive) LD formula for neutrons.
With this, the binding energy of the object becomes:
\begin{eqnarray}
\frac{BE}{N}= a_V-a_A + \frac{3 G m_n^2}{5  r_0} N^{2/3 }\, .
\end{eqnarray}
The gravitational term on the right-hand-side, in contrast to the nuclear terms, is positive and increases with $N$. 
As explained in Part 1~\cite{Part1_2020}, we may find the limits of stability of such an object by looking for the solution of the equation $\frac{BE}{N}=0$. 
For $N$ below a given threshold, $N_0$, the binding energy is negative, indicating that gravity is not strong enough to bind the system. In contrast, for $N_S \ge N_0$, the system is bound and may therefore exist. 
We can easily set up an equation for the threshold value $N_0$:
\begin{eqnarray}
0=
a_{V}-a_A + \frac{3 G N_0^2 m_n^2}{5 N_0 r_0 N_0^{1/3}} =
a_{V}-a_{A} + \frac{3 G  N_0^{2/3 }m_n^2}{5  r_0} \, .
\label{eq:zero}
\end{eqnarray}
\noindent This equation is easily solved analytically. Using the values from Tables~\ref{tab:coeff}  and \ref{tab:conv} for the different constants, we find
\begin{equation}
N_0=\left[ \frac{5 r_0}{3Gm_n^2}(a_{A}-a_V)\right] ^{3/2}\approx 5\cdot 10^{55}\;.
\end{equation}
\noindent Students can work out the numbers of this expression themselves. Alternatively, the tab \verb+Minimum Mass+ in the worksheets provides a direct calculation of $N_S$ for different values of $a_A$ and $a_V$.

From our model, we can therefore expect the existence of a very massive, neutron-rich \emph{nucleus} with $N_S > N_0 \approx 5 \times 10^{55}$ neutrons. The corresponding threshold radius is of the order of $R_0 \approx r_0 N_0^{1/3}\approx4 \text{ km}$. This object would have a mass of $M_S \approx 0.04 M_\odot$, where the solar mass $M_\odot = 2 \cdot 10^{30}$ kg has been introduced. 
We note that the appearance of this mass scale indicates that the macroscopic object we have considered may be of an astronomical nature. 
These values are obtained in our worksheets by finding the solution to Eq.~(\ref{eq:zero}). Because the gravitational term is essentially fixed, students can change the values of $a_V$ and $a_A$, following their work in Part I~\cite{Part1_2020}, to explore how the predictions for $N_S$, $M_S$ and $R_S$ change with the LD formula parameters.

By using this simple model, we have been able to make an \emph{extrapolation} to the LD model. The demonstrator can now challenge the students to think if such an object exists or not. This may bring the discussion towards the concept of \emph{validation}, which is crucial for any scientific model. We encourage demonstrators to address clearly this point to students. A good model is capable of making \emph{predictions}, within or outside its original domain of applicability, that can be validated or invalidated by data.

\begin{figure*}[!t]
\begin{center}
\includegraphics[width=0.6\textwidth,angle=0]{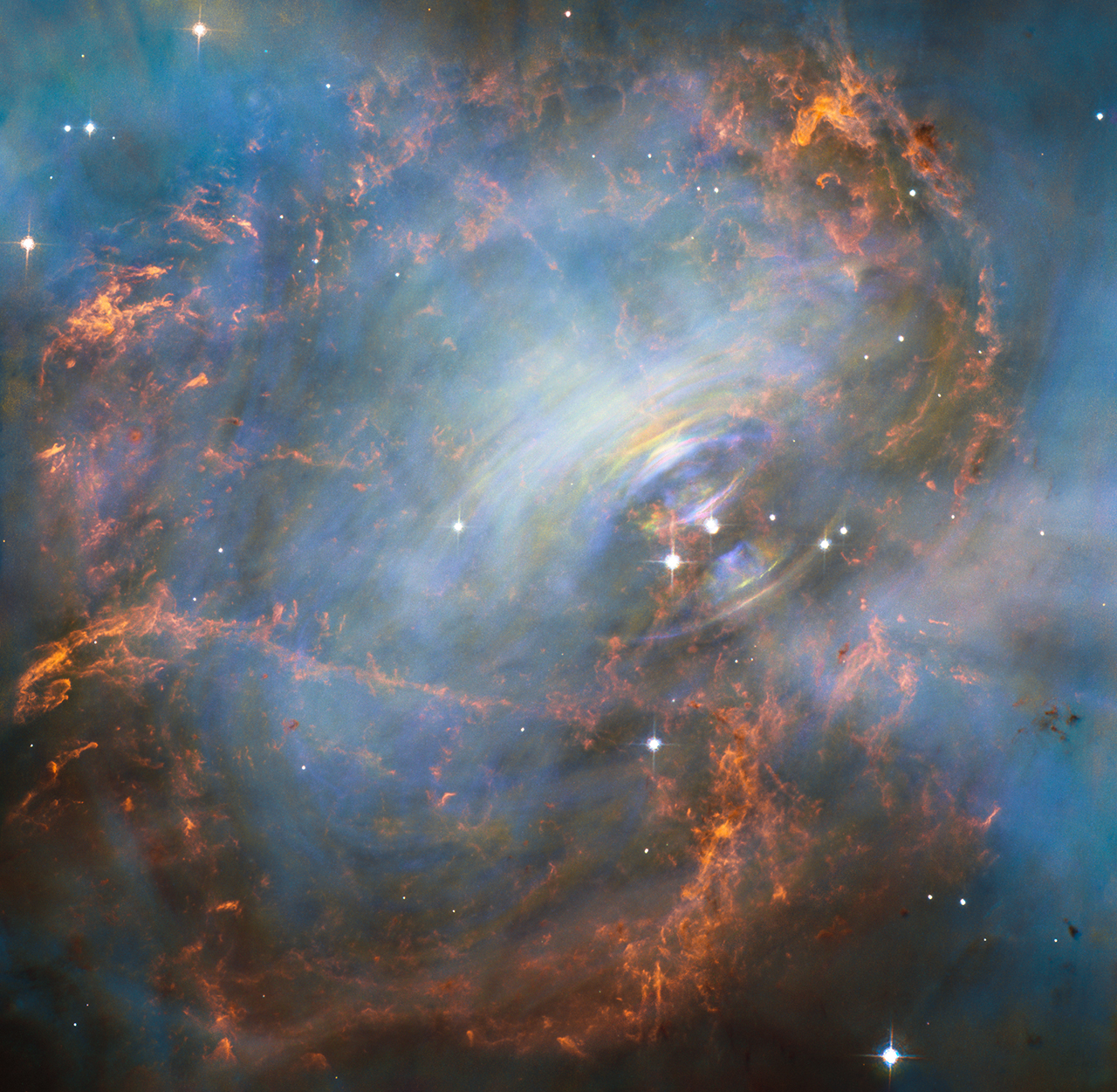}
\end{center}
\caption{(Colour online) The Crab nebula seen by the Hubble telescope. A rapidly spinning neutron star  (the rightmost of the two bright stars near the center of the image) sits at the core of the nebula. In the image, we also observe the bright wisps  generated by the rotation of the neutron star.
Credits: NASA and ESA. Acknowledgment: J. Hester (ASU) and M. Weisskopf (NASA/MSFC)~\cite{nasa}.
}
\label{crab}
\end{figure*}

\subsection{Astrophysical considerations of neutron stars} \label{subsec:astro}

The object we obtained in the previous Subsection is a schematic representation of a real astronomical body: a neutron star (NS)~\footnote{\url{https://www.youtube.com/watch?v=F1Kml3zuTco}}.
A NS is the leftover of a core-collapse supernova which is not massive enough to produce a black-hole remnant. NSs were first conjectured to exist soon after the discovery of the neutron in the 1930s \cite{Landau1932,Baade1934,Yakovlev2013}, but they were only discovered by Bell-Burnell and collaborators as pulsars in the 1960s \cite{hewish1968}. 
Since then, several key steps in the astrophysics of compact objects have allowed for the observation of many more NSs. 

Fig.~\ref{crab} provides an illustration of the Crab nebula, which hosts a central NS.
The Crab nebula played an interesting role in linking radio pulsars, supernovae and NSs together. Demonstrators can use it as an illustration of how some of the ideas around NSs are linked together with stellar evolution. The Crab was known to be the result of a supernova explosion, observed by Chinese astronomers in 1054. Shortly after the first discovery of pulsars, Pacini suggested that the unusually large intrinsic brightness of this nebula could be powered by a rotating neutron star at its center~\cite{Pacini1968}. The discovery of a radio pulsar in the Crab nebula strengthened the credibility of the pulsar-neutron star-supernova association~\cite{Comella1969}. In fact, the fact that pulsars are indeed rotating neutron stars, as first suggested by Gold~\cite{Gold1968}, is nowadays uncontested. 

NSs are the densest stable objects known, and are observed across the electromagnetic spectrum. Inferred magnetic fields at the surface of NSs are possibly the largest in the universe, as are the gravitational fields at their surface. Radio observations of binary systems with at least one NS give access to very accurate measurements of NS masses~\cite{Lattimer2012,Ozel2016}. A recent compilation, with an intuitive figure, can be found in Ref.~\cite{Freire_web}.
Typical masses of NSs in the galaxy are of the order of $1.4 M_\odot$, which is typically taken as a \emph{canonical} value. But NSs as heavy as $2M_\odot$ have also been observed using advance radioastronomy techniques \cite{Demorest2010,Antoniadis2013,Cromartie2020,Cromartie2021}. NS radii are notoriously more difficult to observe, and are expected to be in the range $8-12$ km~\cite{chamel2008physics,Abbott2018}. 
Even more interestingly, the very recent measurements of NS mergers via gravitational waves has kick-started the field of multimessenger astronomy~\cite{abbott2017multi}. We refer to Ref.~\cite{boyle2019teaching} for a simple introduction on the subject of gravitational waves. 

At this stage, the demonstrator may provide a brief overview of the extreme interior properties of NSs. 
In discussing these aspects, demonstrators (or the associated activity online material) should stress the fact that these objects are routinely observed, and that some of their properties are very well known. The NS model in our worksheets predicts the \emph{minimum} mass of a neutron star, and should be taken as a theoretical lower minimum. The order of magnitude result that was obtained in our exercise is close to expectations from more realistic calculations \cite{Haensel}, but is substantially lower than the observed minimum mass (which is of order $\approx 1.1 M_\odot$)~\cite{Lattimer2012,Ozel2016}. This fact can be used to discuss the difference between a theoretical limit and an observational one. For NSs, the latter is likely to be associated to the formation dynamics of NSs, since the generation of supernova shock waves likely requires central objects with masses $\approx 1M_\odot$ \cite{Burrows2020}. Selection effects that could influence our ability to find certain pulsar masses are generally understood~\cite{Kiziltan_2013}. 

Interestingly, the radius $R_S$ predicted by our model is only within a factor of $2$ of quoted literature values~\cite{shapiro1983black,Haensel}. Of course, the application of the LD model to the NS case has several limitations. Given the enormous amount of mass concentrated in a tiny region of space, one should use general relativity and not classical physics to describe the interior and the gravitational energy of this object~\cite{Misner,shapiro1983black}. Quantum mechanical effects are also relevant, but our simple model with only a few parameters has been able to grasp some important physical features of a NS.

%%%%%%%%%%%%%%%%%%%%%%%%%%%%%%%%%%%%
\subsection{Maximum mass}
\label{subsec:max}

At this stage, the students would have worked with the nuclear LD formula to extrapolate the properties of NSs. The instructor could ask the student to find the mass density of the minimum mass object from the previous section, $\rho=\frac{M_S}{V_S}=\frac{M_S}{ \frac{4 \pi}{3} R_S^3} \approx 3 \times 10^{17}$ kg m$^{-3}$ and $V_s$ is the volume of our (spherical) star. This density is comparable to that at the interior of a nucleus. The densest ``terrestrial" materials, like gold, have mass densities of the order of $\rho \approx 2 \times 10^3$ kg m$^{-3}$ - $14$ orders of magnitude smaller. 

At this stage, students can also work out other extreme properties of NSs. One can, for instance, estimate the energy density stored in the neutron star's mass, by using Einstein's mass-energy equivalence fromula, $\epsilon = \frac{E}{V_s} = \frac{M_S c^2}{V_S} = \rho c^2 = 2.7 \times 10^{34}$ J m$^{-3}$. Gasoline, a typical fuel, has an energy density of $\epsilon \approx 3 \times 10^{4}$ J m$^{-3}$. Thus, if all the neutron star mass could be converted into energy, one would generate an immense amount of energy - as observed in neutron-star binaries. 
We stress that these extreme properties can only be accessed through theoretical models, since there are no known experiments that can test the properties of supradense matter at equilibrium. 

A full description of the extreme interior properties of NS requires general relativity. Without entering into cumbersome details, a demonstrator could introduce the need for a general relativistic treatment as follows. 
The Schwarzschild radius of a black hole, which is a prototypical general relativistic object, is given by the equation $R_\text{Sch}= \frac{2 GM}{c^2}$, where $M$ is the mass of the black hole enclosed within this radius \cite{Misner}. For an object of mass $M$ and radius $R$, the compactness parameter,
\begin{equation}
C=\frac{R_\text{Sch}}{R} = \frac{2 G M}{R c^2} \, , \label{eq:compactness}
\end{equation}
measures the deviation with respect to the corresponding Schwarzschild radius. Objects with $C \ll 1$ can safely be consider to be Newtonian, whereas objects with $C \approx 1$ require general relativistic treatments. NSs with typical masses $1.4 M_\odot$ and $R\approx 10$ km have a compactness of $C\approx0.4$ - close to the general relativistic limit. 
 
We do not provide any details of the general relativistic treatment of NS interiors, but rather resort to the solution of the simplest available analytical model~\cite{Misner,Haensel}. This model makes the (unrealistic) assumption that the star is at a constant interior mass density, $\rho_0$. The model is described in detail in in Box 23.2 of Ref.~\cite{Misner} and in Ref.~\cite{Haensel}.
In this oversimplistic picture, the density is the same from the center of the star to its surface and, as a consequence, the mass, $M$, and the radius, $R$, of the star are given by the relation
\begin{equation}
M= \frac{4 \pi}{3} \rho_0 R^3 \, .
\label{eq:MR}
\end{equation}
The general relativistic equations of a spherically symmetric star can be solved analytically and yield the following function pressure profile,
\begin{equation}
p(r) = \rho_0 c^2 \frac{ \sqrt{1-C \frac{r^2}{R^2} } - \sqrt{1-C} }{ 3\sqrt{1-C} - \sqrt{1-C  \frac{r^2}{R^2}} } \, .
\label{eq:profile}
\end{equation}
Here, $r$ is the radial variable within the star, running from $r=0$ at the center through to $r=R$, the star's radius, at the surface. $C$ is the compactness in Eq.~(\ref{eq:compactness}). 
Starting from a central pressure, $p_c$, 
\begin{equation}
p_c=p(r=0) = \rho_0 c^2 \frac{ 1 - \sqrt{1-C} }{ 3\sqrt{1-C} - 1 } \, ,
\label{eq:pc}
\end{equation}
the pressure gradually decreases until it reaches the surface of the star, $r=R$, where $p(r=R)=0$.

\begin{figure*}[t]
\begin{center}
\includegraphics[width=0.7\textwidth]{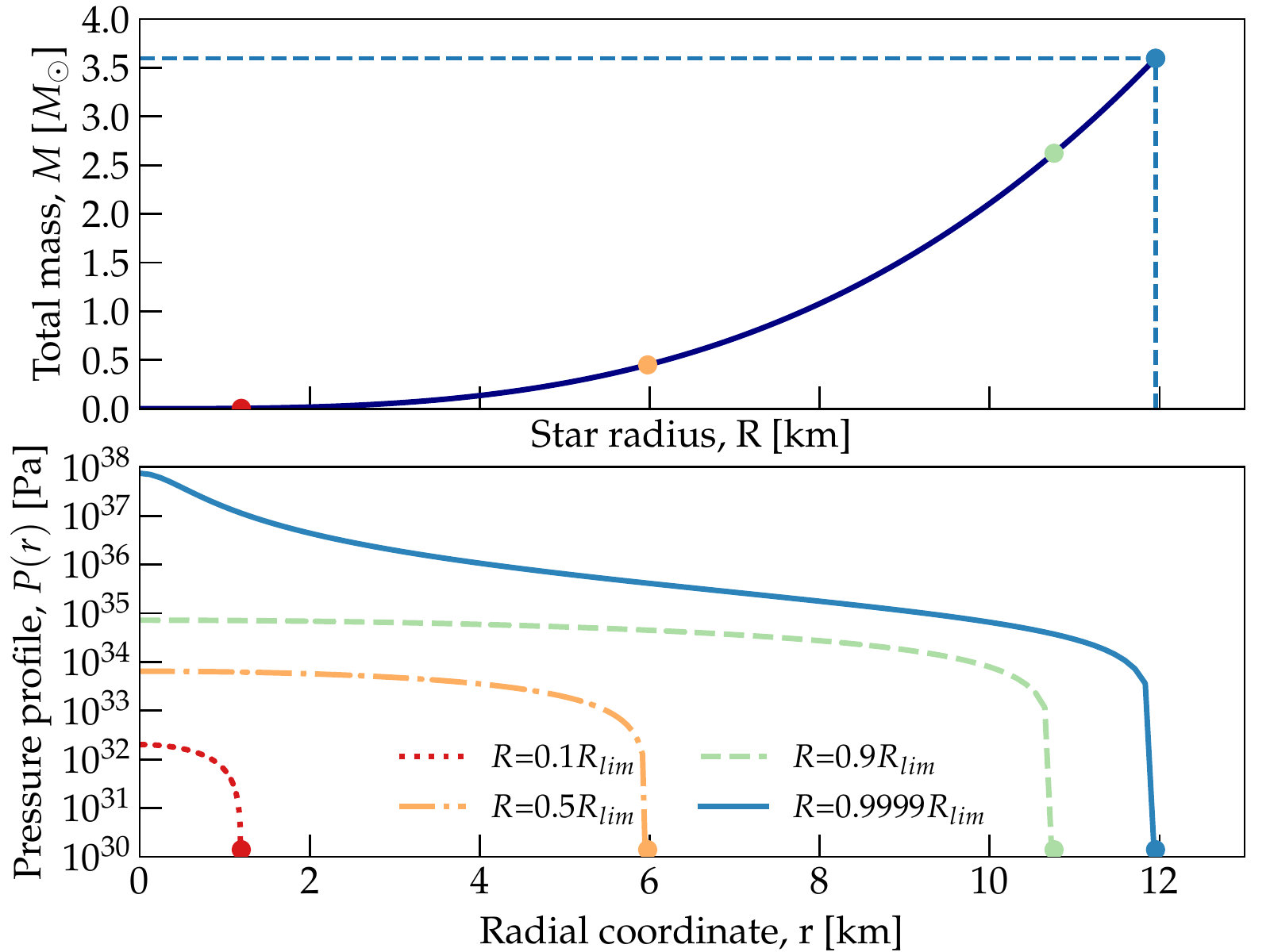}
\end{center}
\caption{(Colour online) Top panel: mass-radius relation of neutron stars obtained with the constant density model with $\rho_0=10^{18}$ kg m$^{-3}$. Bottom panel: interior pressure profiles of neutron stars. The corresponding stellar radii are marked with circles.
}
\label{fig:rho_model}
\end{figure*}

To obtain numerical values for this pressure profile, one requires a value of the interior density, $\rho_0$ and a value of the radius, $R$. The compactness, $C$, can be directly related to $R$ and $\rho_0$ using  Eq.~(\ref{eq:MR}), so that $C=\frac{ 8 \pi G \rho_0}{3 c^2} R^2$. In the tab \verb+Maximum Mass+, students are asked to provide typical values of $\rho_0$ and R. The pressure profile is then immediately calculated and displayed in a plot. Mathematically able students may be provided with Eq.~(\ref{eq:profile}) and asked to work with it as a function - to find where it becomes maximum or minimum. 

For small radii, the compactness will be small and so will the central pressure, $p_c$. As $R$ increases, the denominator in Eq.~(\ref{eq:pc}) approaches zero. Clearly, to avoid the appearance of an unphysical infinite central pressure, we must require that compactness lies below the limiting value $C_\text{lim} < \frac{8}{9} $. In turn, this means that the star cannot have a radius that is larger than $R < R_\text{lim} = \frac{c}{\sqrt{3 \pi \rho_0 G} }$. This also translates into a limit for the highest possible mass given a value of $\rho_0$,
$M < M_\text{lim}=\frac{4 c^3}{\sqrt{3^5 G^3 \pi \rho_0}} $. For a typical value of $\rho_0=10^{18}$ kg m$^{-3}$, these limits correspond to $R_\text{lim} \approx 12$ km and $M_\text{lim}\approx 3.5 M_\odot$. These are qualitatively and quantitatively (within a factor of $2$) close to astrophysical observations of typical radii and maximum NS masses. Again, the instructor may point to Eqs.~(\ref{eq:profile}) or ~(\ref{eq:pc}) to students that want to find the limit themselves, with some very simple mathematical work.

The appearance of this \emph{limiting} mass is very important. It indicates that, for a given value of the central density, the equations of general relativity simply cannot support a star with a mass above $M_\text{lim}$. This is reminiscent to the concept of the Chandrasekhar mass of white dwarfs. Note, however, that the white dwarf limit arises as a consequence of the fact that the electrons providing pressure in a dwarf become relativistic above a certain threshold \cite{Balian1999}. The physical mechanism underlying the Tolman-Oppenheimer-Volkoff limit of neutron stars is different and it is entirely due to general relativity. A constant-density star described in Newtonian physics, like a white dwarf, does not have a limiting mass. 

 An example of the $M-R$ relation (top panel) and the corresponding pressure profiles (bottom panel) for a fixed value of $\rho_0=$ and changing $R$ values are provided in Fig.~\ref{fig:rho_model}. The top panel illustrates the evolution of the mass and radius of the star. More realistic simulations provide an increasing mass as the radius decreases, the opposite behaviour to what is observed in this simple model. In fact, we know that this dependence  is not physical~\cite{Haensel}. Having said that, the model is simple, analytical and provides an intuitive understanding of the interior properties of neutron stars. 
 
\begin{figure*}[!t]
\begin{center}
\includegraphics[width=0.6\textwidth,angle=0]{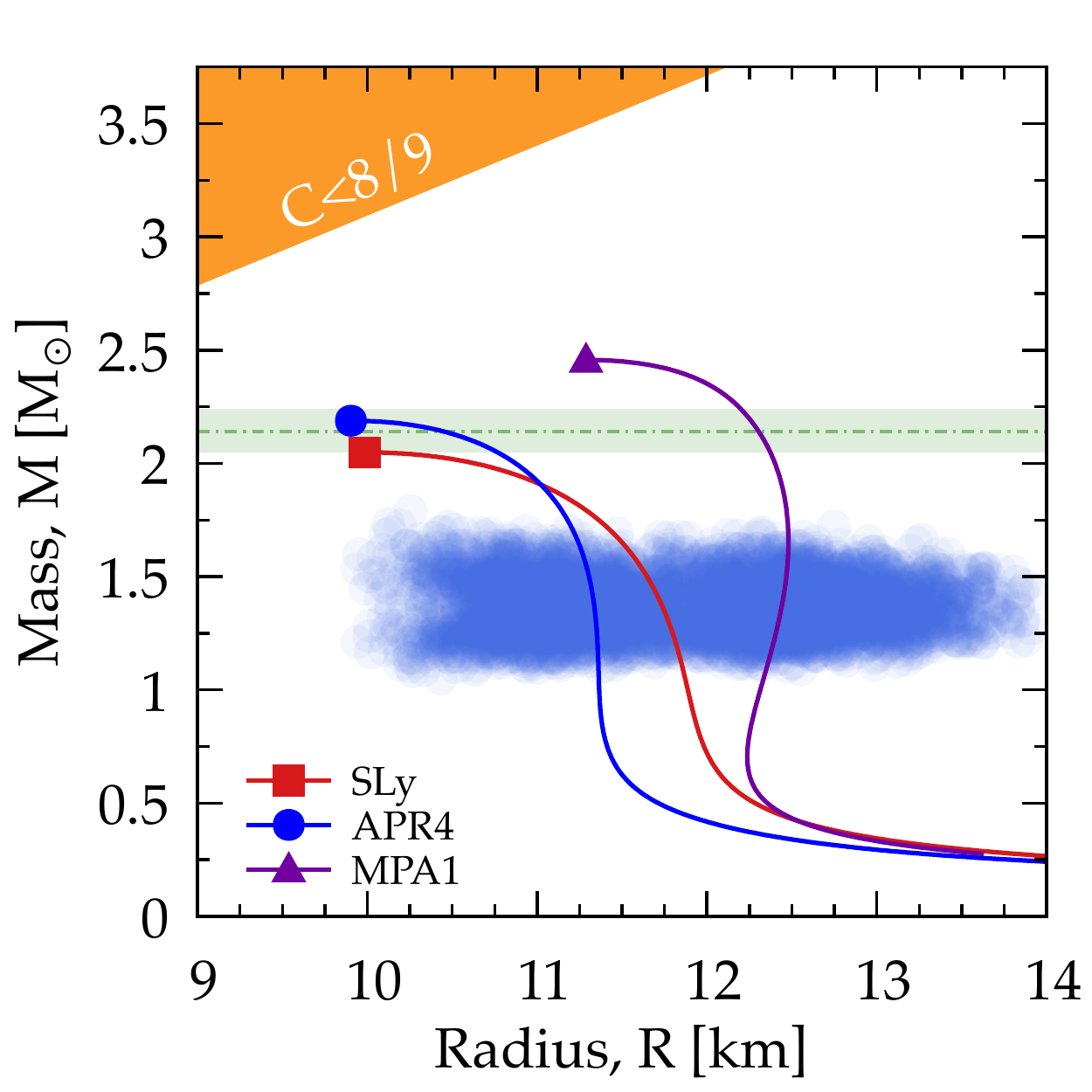}
\end{center}
\caption{(Colour online) Mass-radius relation of neutron stars obtained with three different equations of state: SLy, APR4 and MPA1 (see Ref.~\cite{Read2009} for details). The dot-dashed line correspond to the NS mass with a maximum central value, observed in Ref.~\cite{Cromartie2021}. The associated band denotes the $68 \%$ confidence interval. The blue probability cloud corresponds to the results inferred by the LIGO-VIRO collaboration upon observing GW170817 \cite{Abbott2018}. The orange region delineates the region of maximum compactness predicted by the constant density model.   
}
\label{fig:MR}
\end{figure*}

 In more realistic simulations of NS structure, the mass-radius relationship depends on the equation of state of dense matter, which related pressure to energy density in the star's interior. The latter can be computed with a variety of theoretical techniques, but several uncertainties remain \cite{Lattimer2012,Ozel2016}. Different equations of state predict different mass-radius relationships and, also, different limiting masses~\cite{lattimer2011two}. This means that a single measurement of a new maximum neutron-star mass can single-handedly invalidate several theoretical predictions. We present typical mass-radius relationships for three different equations of state in Fig.~\ref{fig:MR}. The most recent and largest (central value) neutron star mass observed to date in Ref.~\cite{Cromartie2020} (recently updated in Ref.~\cite{Cromartie2021}) is displayed in a horizontal dashed-dotted line, together with a confidence interval. If this interval was narrower, it may invalidate one of the interior neutron-star models, the well-known SLy equation of state~\cite{Haensel}. At the time of submitting this paper, the NICER collaboration, based on an X-ray telescope onboard the International Space Station, had announced (but not yet published) radius measurements for this NS. 
 
 The top left region of Fig.~\ref{fig:MR}  shows the region in the mass-radius diagram that is excluded by the compactness limit of the constant density model discussed earlier. While this is not more constraining than current measurements or theoretical predictions, it illustrates the usefulness of this model in a simple research setting.
 
In the tab \verb+Maximum Mass+, students have a notebook that displays the pressure profile starting from a value of the central density, $\rho_0$, and star radius, $R$. This allows for a clear understanding of the orders of magnitude of pressure within a star. The instructors can use this visual display to indicate the exotic properties of superdense matter. For instance, typical central pressures of NSs are of the order $P_c \approx 10^{34-38}$ kg m$^{-3}$. The pressure at the bottom of the Marianna trench is, in contrast, only $P \approx 10^8$ Pa, whereas the central pressure on Earth is $P \approx 3 \times 10^{11}$ Pa. These are many orders of magnitude smaller than NS values.

In addition, the tab provides the corresponding mass of the neutron star with a given radius. By increasing the value of the radius $R$, the students may go above the value of $R_\text{lim}$. This is produces no pressure profile in the plot. Demonstrators explaining this can then request the students to play with the spreadsheet, until they find a relatively accurate value of $R_\text{lim}$ and the corresponding $M_\text{lim}$. One way of doing this would be to check for values of $R$ that provide steeply increasing central pressure profiles, as observed in the solid line of Fig.~\ref{fig:rho_model}. 

There is another aspect that can be discussed in the context of this work - and that is the idea that models can predict their own \emph{limits}. This is a non-trivial ideal. As we have just seen, scientific models can be extrapolated and, if taken too far, may provide incorrect results. This is a model \emph{limitation}, and its origin lies in the extrapolation procedure. In contrast, the general relativistic constant density model indicates a real, physical limit to the system itself. Not all models can predict limits ``on their own", and the constant-density model is  possibly one of the simplest exceptions. A related limit in general relativity is that of a black hole singularity \cite{Misner}. 

\section{Conclusions}\label{sec:concl}

In this article, we have discussed in detail the concept of \emph{extrapolation} using a simple liquid drop model to study basic properties of massive astrophysical objects. A key advantage of theoretical and mathematical modelling is the ability to look into  regimes that cannot be necessarily explored by experiments. We illustrate this point here by providing two activities that look into the basic features of massive neutron stars. The first activity is a clear continuation of the LD model as discussed in detail in Part 1~\cite{Part1_2020}. By using the worksheets provided as Supplementary Material, the student will be able to judge the dependence of the extrapolation of the results on the input parameter of the simulation. The latter have been adjusted in the activity in Part 1.

In contrast, the second activity is purely astrophysical in nature and provides a direct exploration of the interior properties of NSs.
A major outcome of these activities is to provide a different view on nuclear physics. In particular, the application to astrophysical objects indicates the relevance of nuclear physics across domains, not only on applications, such as energy production or nuclear medicine, but also on the critical impact of the understanding of important phenomena taking place in our Universe.  The activities allow students to familiarise themselves with one of the most exotic compact objects in nature, and provide an understanding of the orders of magnitude involved in terms of mass density and pressure. 

The activities provided here can also be used as a starting point for a deeper discussion into scientific models. Following the validation and the prediction steps discussed in Part 1, these activities illustrate the extrapolation process. While we focus here on a successful application of extrapolation, it is important to discuss model limitations as well. In fact, the second activity clearly points out to a limit in the model itself. It may be important to point out that such limits are rare in physics and have important conceptual implications.

\ack
The \emph{Binding Blocks} project has been funded by EPSRC and University of York through an EPSRC Impact Acceleration Award, and by an STFC Public Engagement Small Award ST/N005694/1 and ST/P006213/1. The work of ARH is funded through grant ST/P005314/1; and by the Spanish State Agency for Research of the Spanish Ministry of Science and Innovation through 
the ``Ram\'on y Cajal" program with grant RYC2018-026072 and
the ``Unit of Excellence Mar\'ia de Maeztu 2020-2023" award to the Institute of Cosmos Sciences (CEX2019-000918-M). We thank the PHAROS COST Action (CA16214) for partial support.

\section*{References}
\bibliographystyle{iopart-num}
\bibliography{biblio}

\end{document}